\documentclass[envcountsame,runningheads,notitlepage]{llncs}

\usepackage{cite}
\usepackage{amsmath,amssymb,amsfonts}
\usepackage{algorithmic}
\usepackage{graphicx}
\usepackage{textcomp}
\usepackage{xcolor}

\begin{document}


\title{Electromagnetic Side-Channel Analysis of PRESENT Lightweight Cipher}

\authorrunning{Gunathilake, Lo, Buchanan and Al-Dubai}
\titlerunning{Side-Channel Analysis of PRESENT }

\author{%
Nilupulee A Gunathilake\inst{1}  \and 
Owen Lo\inst{1} \and
William J Buchanan\inst{1} \and
Ahmed Al-Dubai\inst{1}
}

\institute{Blockpass ID Lab, Edinburgh Napier University, UK}

\maketitle


\begin{abstract}


Side-channel vulnerabilities pose an increasing threat to cryptographically protected devices. Consequently, it is crucial to observe information leakages through physical parameters such as power consumption and electromagnetic (EM) radiation to reduce susceptibility during interactions with cryptographic functions. EM side-channel attacks are becoming more prevalent. PRESENT is a promising lightweight cryptographic algorithm expected to be incorporated into Internet-of-Things (IoT) devices in the future. This research investigates the EM side-channel robustness of PRESENT using a correlation attack model. This work extends our previous Correlation EM Analysis (CEMA) of PRESENT with improved results. The attack targets the Substitution box (S-box) and can retrieve 8 bytes of the 10-byte encryption key with a minimum of 256 EM waveforms. This paper presents the process of EM attack modelling, encompassing both simple and correlation attacks, followed by a critical analysis.
\end{abstract}




\section{Introduction}


Side-channel attacks represent significant practical vulnerabilities in cryptographic systems. These attacks exploit secret information, such as encryption keys, by analysing external physical parameters of a device, such as power variations, electromagnetic (EM) emissions and acoustic changes. This phenomenon does not involve internal parameters like the cryptographic algorithm or the code running inside. Analysing these side-channel vulnerabilities, known as side-channel analysis or physical security analysis, falls under practical cryptanalysis.

Performing side-channel attacks was costly and time-consuming during the mechanical era. For example, breaking the Enigma, which was a cipher device used by the German military during World War II in the mid-1940s, required substantial financial resources and several months to achieve success. In contrast, the digital era has made side-channel attacks more affordable and accessible, with antennas to measure EM radiation costing less than £10. Additionally, the increased processing speeds of modern computers allow for more efficient results. Therefore, continual explorations of such side-channel leakages are vital to minimise risks associated with cryptographic operations.

Although EM attacks were not a major concern in the past, it has been identified that EM attacks can cause a significant impact on digital forensics now. Consequently, EM Analysis (EMA) has gained growing attention in academia. However, EMA regarding lightweight ciphers still needs to be improved.

PRESENT is an ultra-lightweight symmetric block cipher introduced in 2007 \cite{PRESENT}, and approved by the International Organisation of Standards (ISO)/International Electrotechnical Commission (IEC) \cite{ISOIEC29167}. The National Institute of Standards and Technology (NIST) has also included it in their report on lightweight cryptography, NISTIR 8114 \cite{NISTIR8114}. PRESENT features a Substitution-Permutation Network (SPN) and a 64-bit block size. It offers two key versions: 80-bit and 128-bit. For lightweight encryption, the 80-bit version is recommended. The algorithm computes through 31 rounds. The Substitution box (S-box) is 4-bit to 4-bit, with corresponding values listed in Table \ref{table:sbox}. Overall, this cipher provides moderate security.

\begin{table*}
\centering
\caption{S-box of PRESENT block cipher (in hexadecimal)}
\label{table:sbox}
\begin{tabular}{|c|c|c|c|c|c|c|c|c|c|c|c|c|c|c|c|c|}
\hline
x & 0 & 1 & 2 & 3 & 4 & 5 & 6 & 7 & 8 & 9 & A & B & C & D & E & F \\
\hline
S(x) & C & 5 & 6 &	B &	9 &	0 &	A &	D &	3 &	E &	F &	8 &	4 &	7 &	1 &	2 \\
\hline
\end{tabular}
\end{table*}

\subsection{Our Contribution}


This study targets the S-box of PRESENT for a Correlation EM Analysis (CEMA) regarding firmware robustness. The choice of an S-box to attack is due to its non-linearity property\footnote{Non-linear properties cause unpredictability in cryptographic functions because the values produced are not sequential}. The 80-bit key version is selected to facilitate lightweight cryptanalysis for the Internet-of-Things (IoT) platform. According to the existing literature, this is the first-ever CEMA study on PRESENT.

This paper presents a well-improved version of our initial work \cite{nilu22}. Our previous work was able to derive up to 7 bytes of the encryption key, and this work is able to retrieve 8 bytes. Our analyses
evaluate the performance against a CEMA of PRESENT supported by a Simple EMA (SEMA) and a Simple EM Frequency Analysis (SEMFA). The analysis is based on a non-invasive white-box EM attack, but also discusses its practicalities in black-box attacks. The attack uses 256 EM waveforms for a successful attack (2048 for optimisation), which is the minimum in known research history for EM attacks. This significantly reduces processing time, as other studies, such as \cite{DEMA_PRESENT} and \cite{cema_twine} have used 12,000 to 17,000 and 15,000 waveforms, respectively.


\section{Electromagnetic Side-Channel Analysis} \label{sec:EMA}


All powered-up electronic devices emit EM radiation generated as a result of the electric current flow in their internal components. In EMA, EM emanation around a device is analysed using mathematical attack models. The aim is to observe whether secret cryptographic keys leak through the excess EM radiation produced during the encryption process. Since real-time scenarios often lack prerequisite knowledge of the encryption process (e.g., which algorithm, which part of the algorithm, how many rounds, etc.), it becomes difficult for an attacker to successfully break into an encrypted data system. Traditionally, oscilloscopes have been used to monitor, investigate, and collect EM waveforms. However, a low-cost alternative, Software-Defined Radio (SDR), has recently emerged in the field. The main mathematical models in EMA are SEMA, Differential EMA (DEMA) and CEMA.

\begin{figure}
        \center{\includegraphics[width=1.0\textwidth]
        {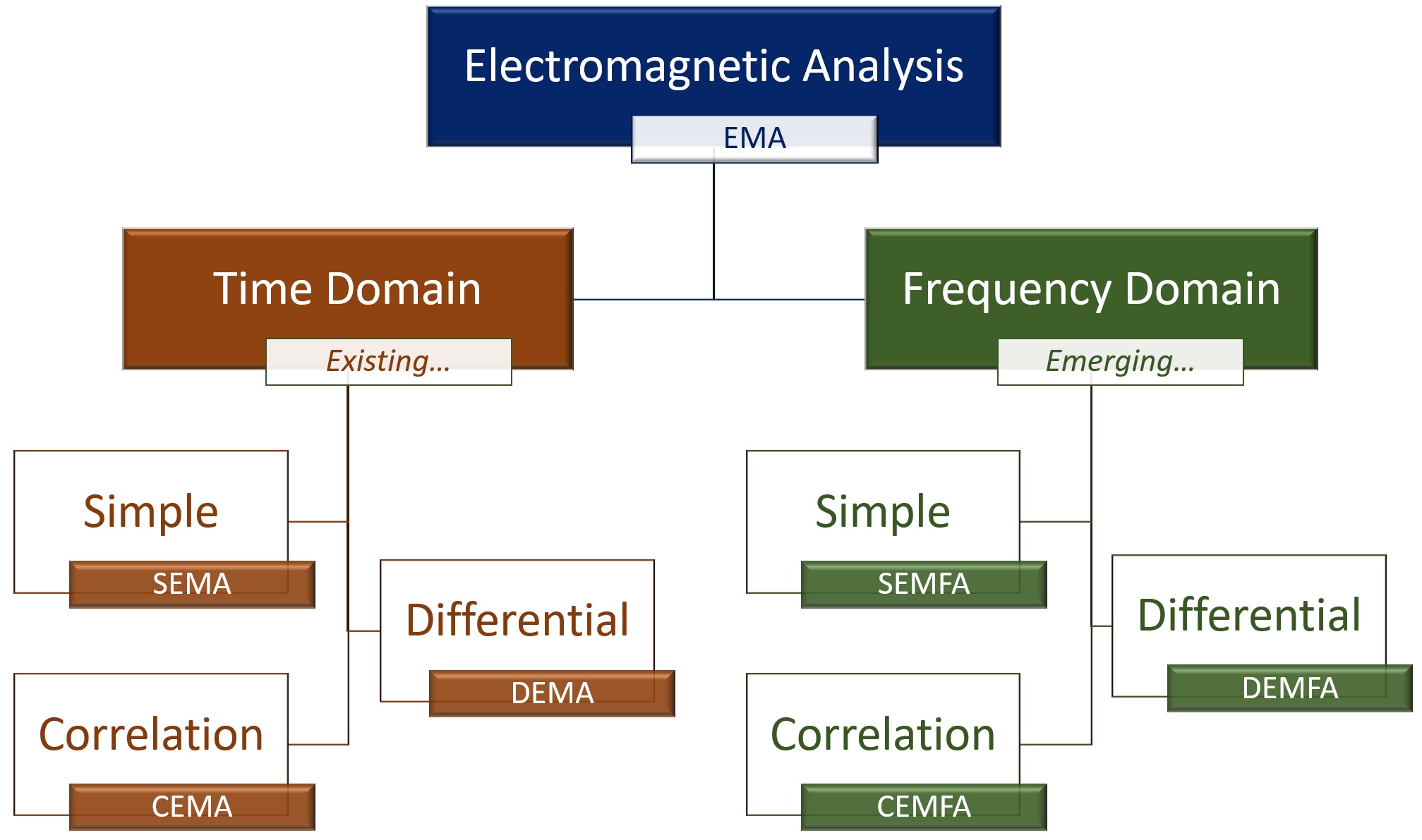}}
        \caption{\label{fig:EMA}EMA classification}
\end{figure}

Although the typical domain for conducting such attacks is the time domain, there has been a growing interest in performing these analyses in the frequency domain recently \cite{EM_freq2,SCA_freq_IoT}. According to \cite{F_SCA_AES}, work in the frequency domain tends to increase the accuracy of outcomes by mitigating trace alignment issues faced in the time domain. The accuracy of alignment is generally sensitive in types of differential analysis, whereas correlation analysis is not greatly affected by this phenomenon. \cite{SEMA} demonstrates that the frequency domain can be utilised to suppress unwanted noise in signals. Whatsoever, a well-thought-out methodology is required because simply converting domains via Fast Fourier Transform (FFT) does not establish a precise attack model. The literature review on EMA concludes the classification of EMA as shown in Fig. \ref{fig:EMA}. When analyses are conducted in the frequency domain, the names of the analyses, SEMA, DEMA and CEMA are adapted to SEMFA, DEMFA and CEMFA. Depending on the attack design, whether it is active or passive, invasive or non-invasive—additional methods such as template attacks \cite{simple_tem_EMA,TemplateSCA}, timing attacks, fault injection, collision and Trojan attacks \cite{SEMA} can be performed in combination.

\subsection{SEMA and SEMFA}


Simple analyses involve direct visual inspections of EM waveforms. It may help identify where leakage occurs if any probability of secured information leakage exists. It usually does not involve retrieving secret data such as private keys, to breach secured data. In contrast, locating the exact cryptographic functionalities and guessing secret information via the device's clock cycles followed by Hamming Weight (HW) changes are possible if rigorous knowledge of the device is available \cite{SEMA,simple_tem_EMA,Sayakkara_2019}. HW can be calculated using Equation \ref{eq:HW}. Invasive attacks are more likely to be successful in retrieving confidential data through simple analyses.

\begin{equation} \label{eq:HW}
    E = g \cdot HW(I) + n
    \end{equation}

    where,
    
    E - Hypothesised EM emission energy
    
    I - Intermediate value
    
    g - Gain
    
    n – Noise


\subsection{CEMA and CEMFA}


In correlation models, knowledge of device specifications or operations is not essential to derive secret information. The process considers several bits at a time, rather than one by one, making it more efficient compared to differential analysis. The major concern in these types of analyses is that the correlation between a hypothesized intermediate value (Equation \ref{eq:HW}) and actual EM data values should indicate the possibility of secret information leakage via the highest correlation magnitudes. The correlation coefficient, which indicates the statistical relationship between two continuous variables, is calculated for this purpose. Pearson’s correlation coefficient\footnote{Named after Karl Pearson}, shown in Equation \ref{eq:pearson}, is widely used for correlation side-channel attacks. It measures the linear correlation between two data sets. Additionally, Spearman's and Gini's coefficients have also been considered in some studies \cite{localised_gini}.

\begin{equation} \label{eq:pearson}
\rho = \frac{Cov(X,Y)}{\sigma_X \sigma_Y}
\end{equation}

where,

\textit{$\rho$ - Pearson correlation coefficient}

\textit{Cov(X,Y) - Covariance between X and Y (equation \ref{eq:covariance})}

\textit{$\sigma_X$ - Standard deviation of X}

\textit{$\sigma_Y$ - Standard deviation of Y}

\begin{equation} \label{eq:covariance}
Cov(X,Y) = E[(X-\mu_X)(Y-\mu_Y)]
\end{equation}

where,

\textit{E - Expectation}

\textit{$\mu_X$ - Mean of X}

\textit{$\mu_Y$ - Mean of Y}


\section{Methodology}

\begin{figure}
        \center{\includegraphics[width=1.0\textwidth]
        {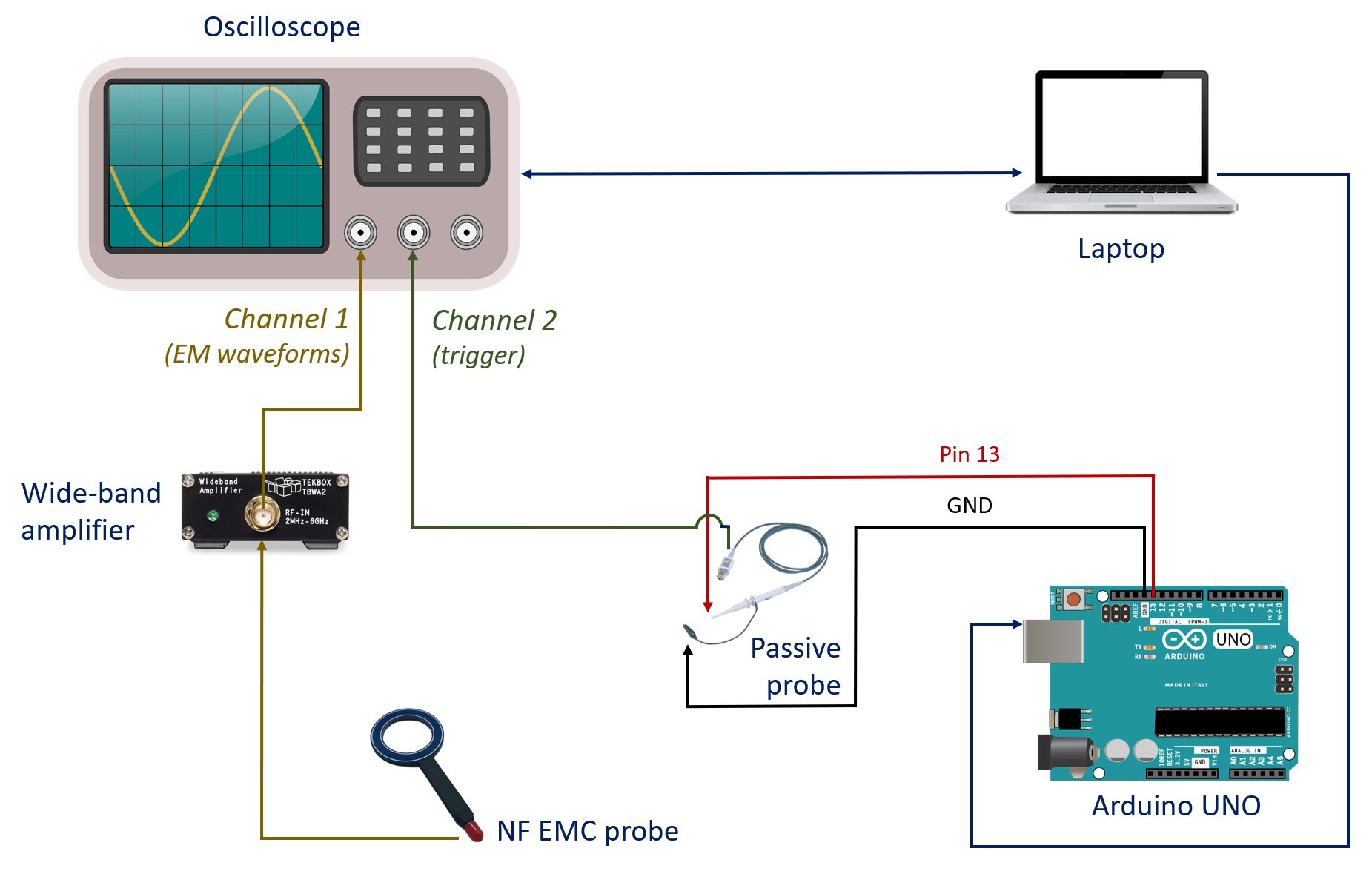}}
        \caption{\label{fig:net}Hardware connectivity of the testbed}
\end{figure}

Since no other research on CEMA of PRESENT was available, predicting the nature of the outcome was quite challenging. Therefore, the experiments were initially conducted using simple analysis and later enhanced to correlation analysis. Fig. \ref{fig:net} shows the testbed used to capture EM waveforms using Near Field (NF) EM Compatibility (EMC) probes. Our method did not use any specifically designed side-channel evaluation hardware or software.

The diameters of the magnetic probe loops are 5~mm, 10~mm, and 20~mm for H5, H10, and H20, respectively. The waveforms were captured through an oscilloscope. The samples were saved on a computer for postprocessing for the analysis. A sample consisted of 256 waveforms for a basic attack, and 2,048 waveforms for an optimised attack. The maximum sampling rate of the oscilloscope was 5~GSa/s. Since two channels were used for the setup, the maximum sampling rate for EM waveforms became 2.5~GSa/s. In the experiments, two probe positions were examined: one 'in parallel' and the other 'perpendicular' to the chip. Finally, the Success Rates (SRs) were calculated changing the bit order of the key in four patterns such that all zeros (00000000b), zero and ones in every other bit (01010101b and 10101010b) and all ones (11111111b) in the key. More information on the setup and methodology is available in \cite{nilu22}.


\section{Results \& Observations}


During encryption, the time taken for the completion of the S-box was roughly 12.73~$\mu$s. The completion of a sample took approximately 20 minutes for 256 waveforms and 2.5~hours for 2,048 waveforms.


\subsection{SEMA}

\begin{figure}
        \center{\includegraphics[width=1.0\textwidth]
        {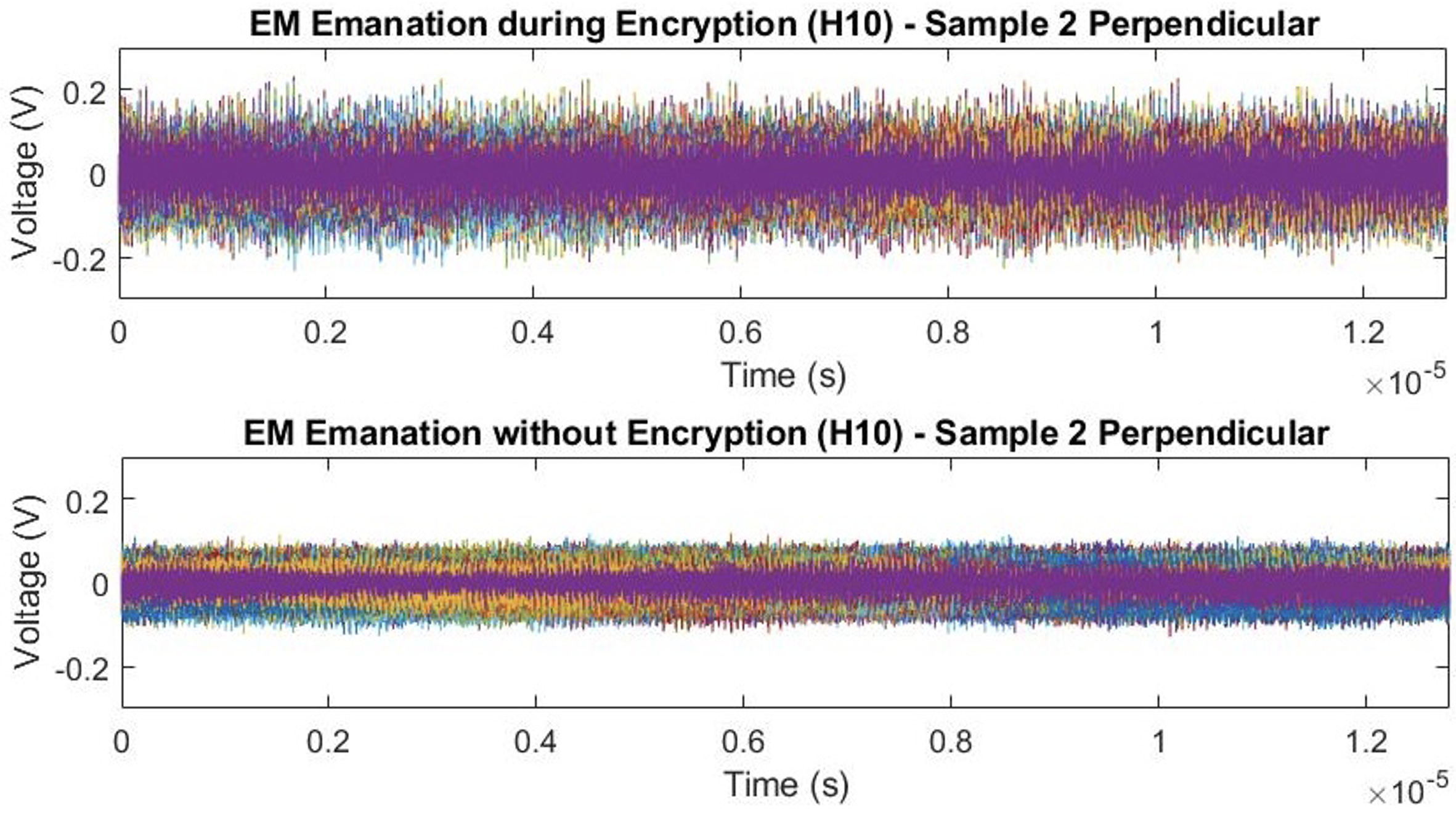}}
        \caption{\label{fig:time}EM emanation differences in time domain}
\end{figure}

The EM radiation of the device nearly doubles when the encryption runs (Fig. \ref{fig:time}). The perpendicular position appears to offer better stability in terms of Radio Frequency (RF) noise avoidance. The induced voltage increases as the diameter of the probe loop increases.

\subsection{SEMFA}

\begin{figure}
        \center{\includegraphics[width=1.0\textwidth]
        {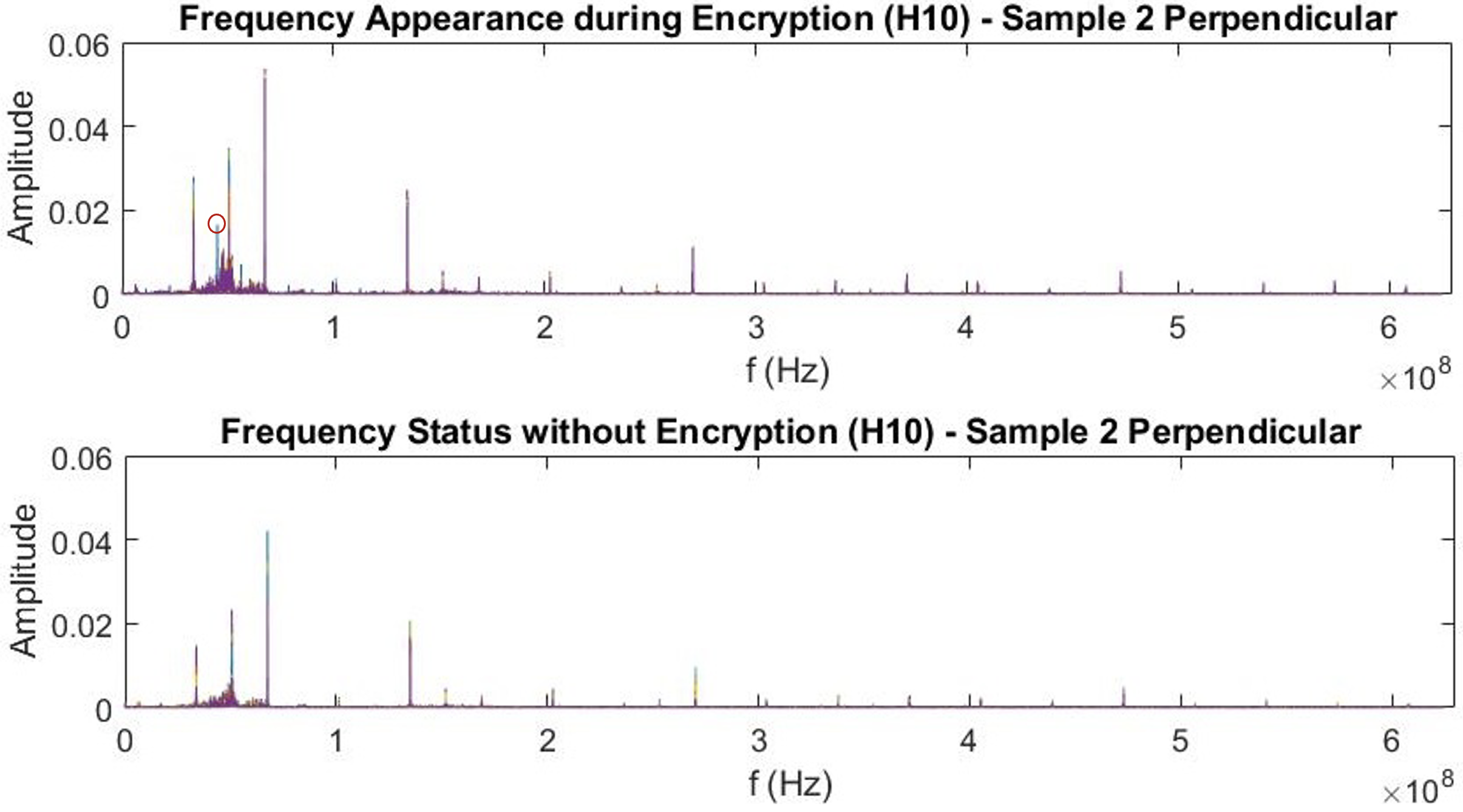}}
        \caption{\label{fig:FFT}EM emanation differences in frequency domain}
\end{figure}
\begin{figure}
        \center{\includegraphics[width=1.0\textwidth]
        {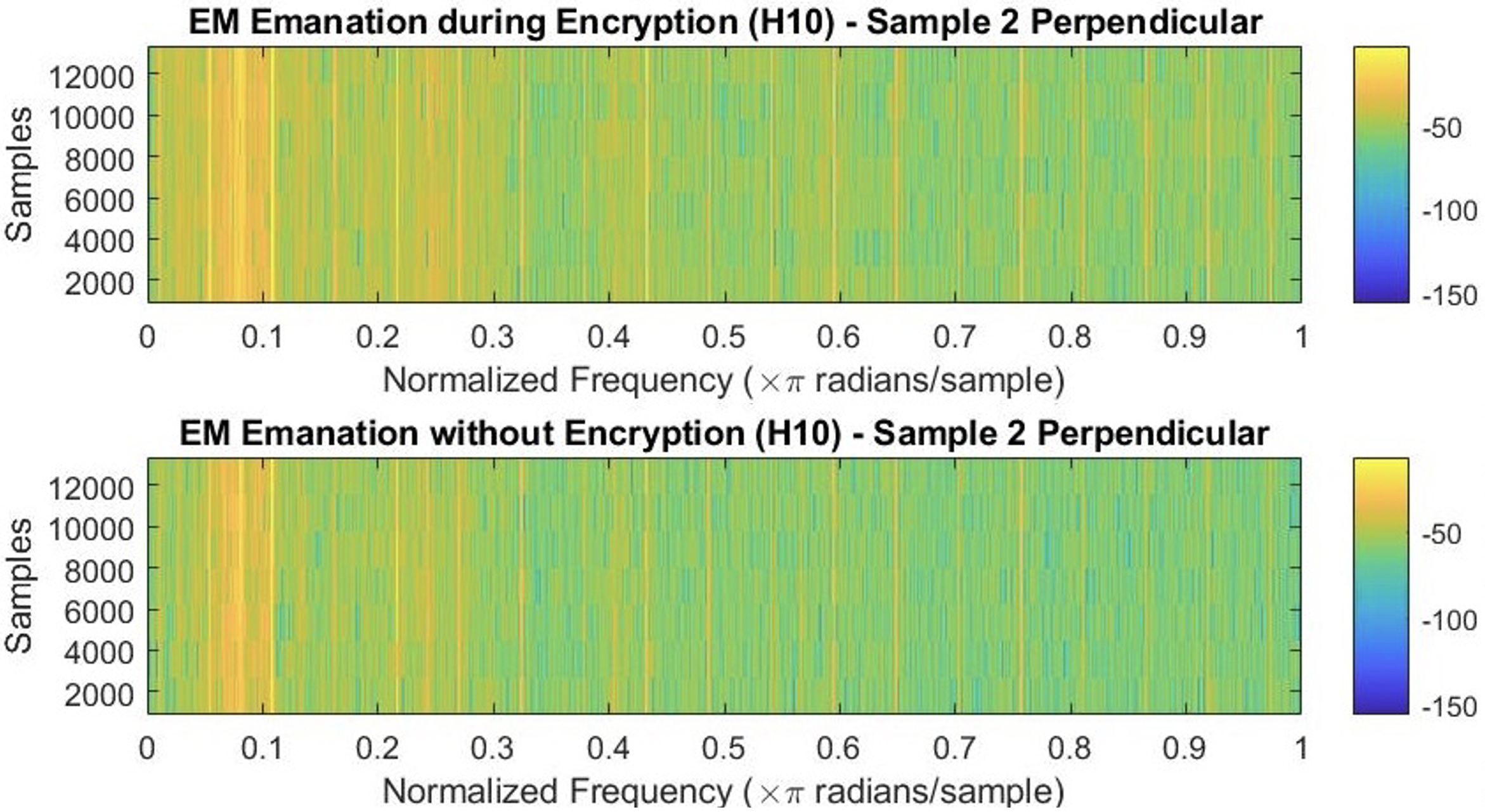}}
        \caption{\label{fig:spectro}EM emanation differences in spectrogram}
\end{figure}

The frequency ranges from approximately 0 to 625~MHz, with significant amplitudes at specific points in the spectrum, notably at 11.25~MHz, 22.5~MHz, 45.08~MHz, 56.33~MHz, 78.83~MHz, 90.08~MHz and 112.66~MHz. The signal strength appears to vary regardless of the probe type and its placement position when the encryption runs. The frequency domain signal and spectrogram differences are in Fig. \ref{fig:FFT} and Figure \ref{fig:spectro}, respectively.

\subsection{CEMA}

\begin{figure}
        \center{\includegraphics[width=1.0\textwidth]
        {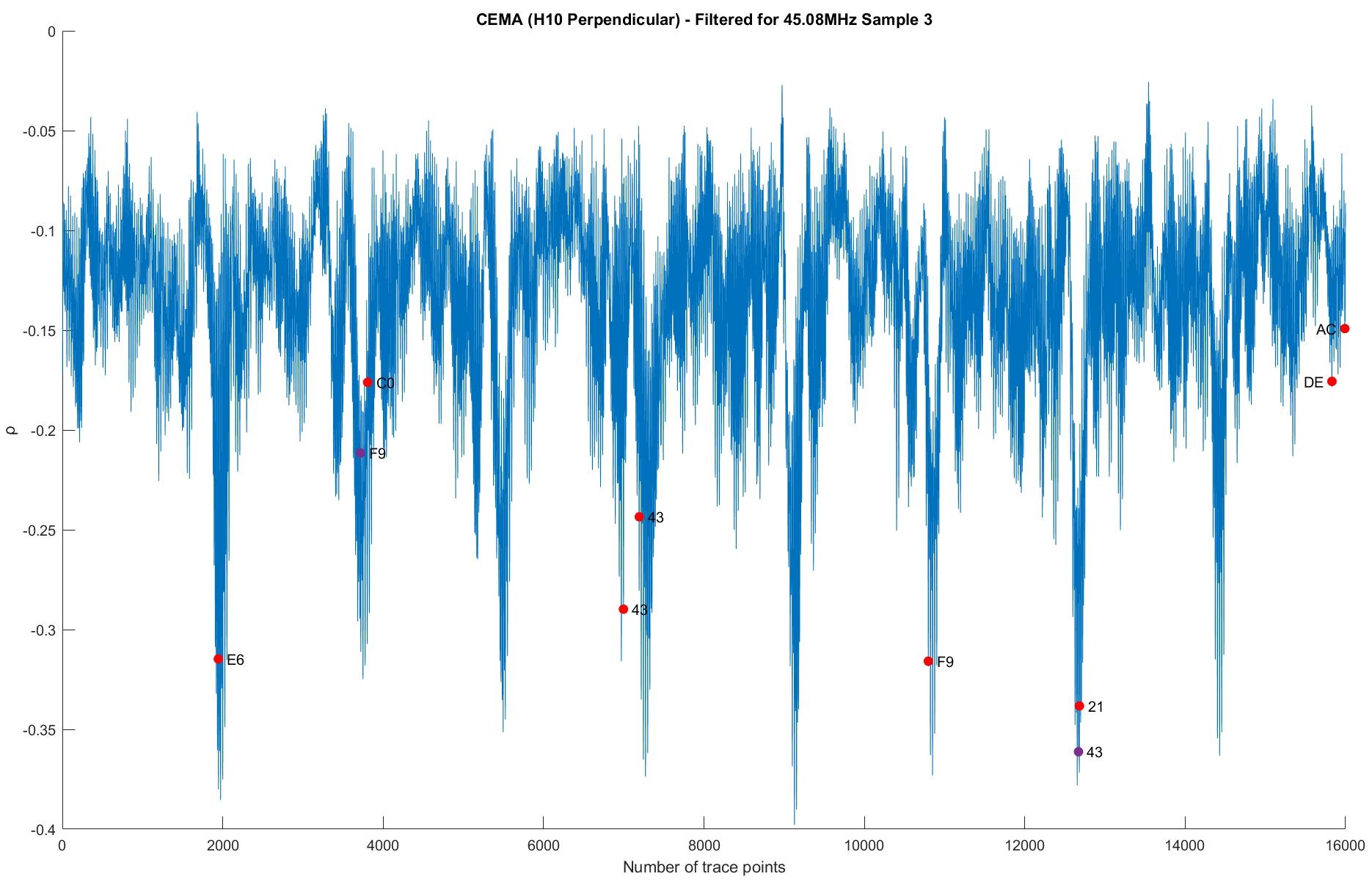}}
        \caption{\label{fig:cema45}Correlation EM attack}
\end{figure}

The attack was completed in less than 20 minutes when the frequencies observed in SEMFA were filtered before post-processing. A reduction in amplitudes was noted for the filtered sample data. It helped in gaining higher correlation magnitudes. Observations indicate that the most suitable NF EMC probe for the task is the H10. A resultant correlation attack graph is in Fig. \ref{fig:cema45}.

\subsubsection{SR}

100\% SR was achieved by the H5 for the 1st byte of 00000000b; the H10 for the 8th and the H20 for the 3rd bytes of 01010101b; and the H5 for the 2nd, the 4th, the 6th bytes, the H10 for the 1st byte, and the H20 for the 2nd byte of 11111111b. Consequently, the SR appears to be higher when the bits of the key are ones instead of zeros.

\subsubsection{Optimisation}

\begin{figure}[!htb]
        \center{\includegraphics[width=1.0\textwidth]
        {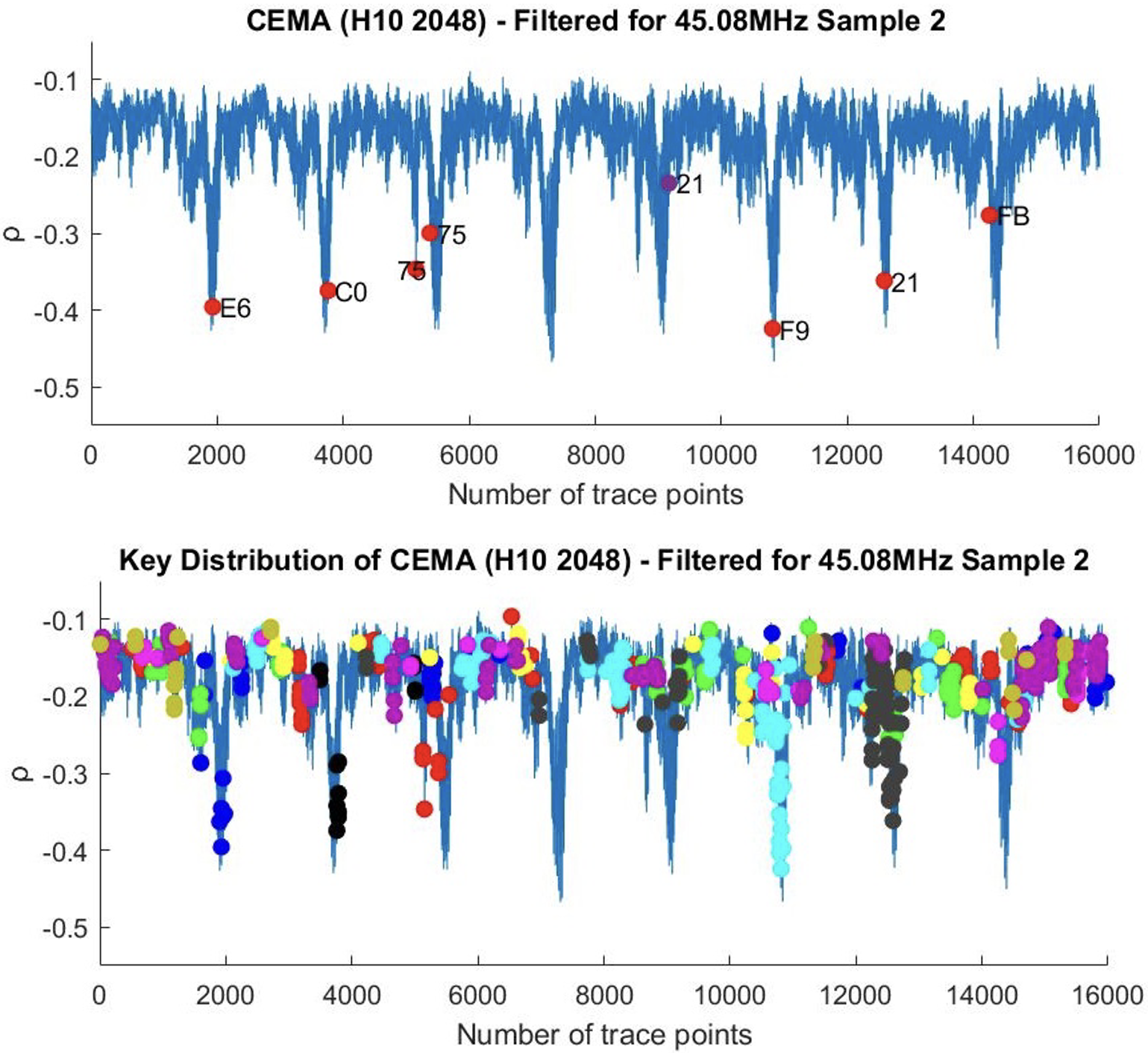}}
        \caption{\label{fig:opt}Attack optimisation results (2048 traces per attack)}
\end{figure}

Filtering and post-processing each sample took more than 2 hours. Compared to the 256 traces per attack, this approach resulted in more stable, improved and sharper correlation graphs (Fig. \ref{fig:opt}). Eight potential leakage areas were identified, with no probability of leakage for the 9th and 10th bytes.

\section{Discussion}

\begin{figure}[!htb]
        \center{\includegraphics[width=1.0\textwidth]
        {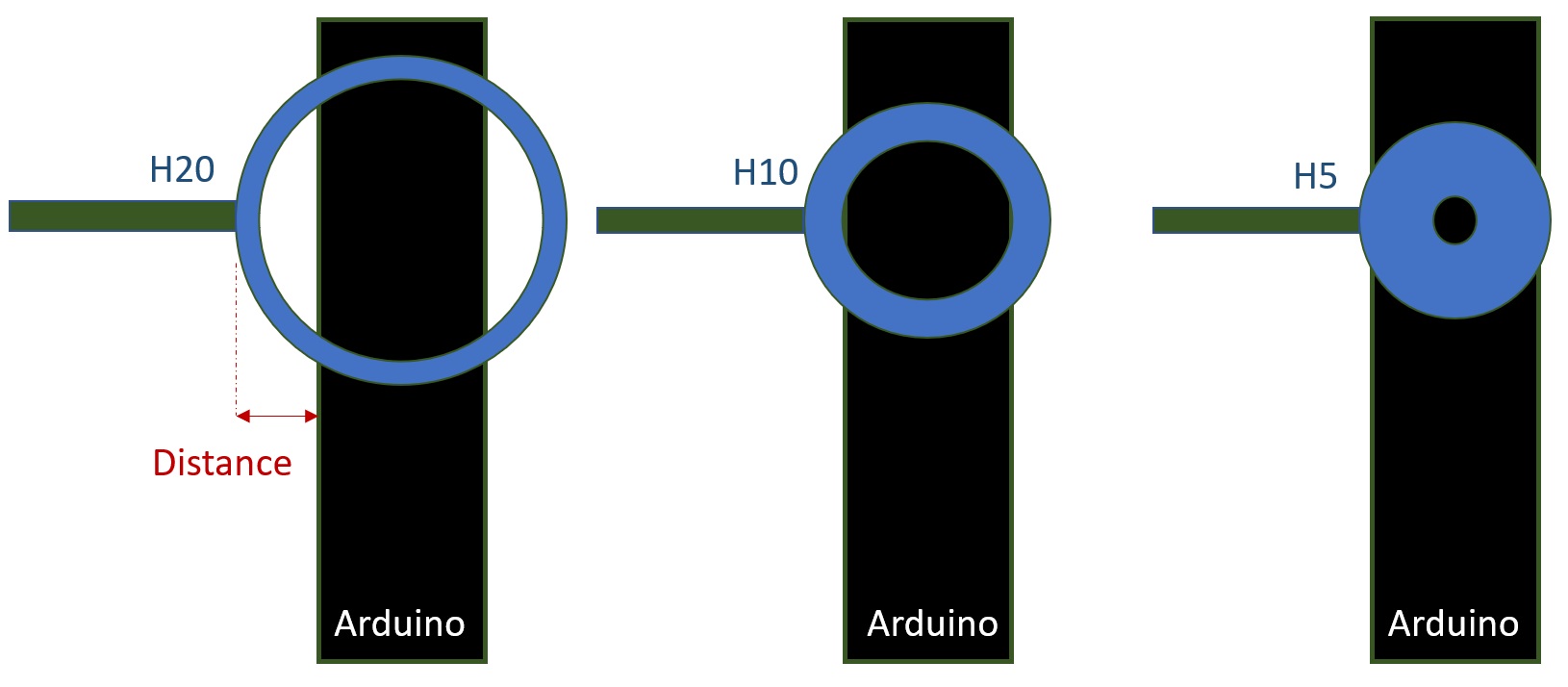}}
        \caption{\label{fig:dia}Area coverage of Arduino UNO chip for different H probe diameters}
\end{figure}

SEMA results confirm that the voltage nearly doubles as a result of the encryption. It appears that the induced voltage increases with the diameter of the probe. Therefore, it can be inferred that a probe with a larger diameter may offer higher correlation magnitudes in correlation attacks. This also depends on the extent to which the probe covers the area of the device chip, as it needs to be sensitive enough to capture the frequencies significantly affected by the encryption. Although \cite{robyns} states that the optimal EM radiation of the Arduino UNO chip can be recorded when the probe is placed between the GND and $V_{cc}$ pins, the loop of the H20 probe seems to be far away from that point compared to other probes as illustrated in Fig. \ref{fig:dia}. Probes with smaller loops have better frequency resolution but are less sensitive. Therefore, compromises in overall outcomes are expected depending on the specific probe type.

The most important information obtained from the SEMFA process is the identification of new frequency components that emerged as a result of the encryption. Differences in EM radiation between the encryption and non-encryption statuses were not visible to the naked eye. Therefore, simple analysis can help predict the likelihood of success for a complex attack model.

According to the results, eight potential leakage areas are clearly noticeable for all magnetic probes even without filtering. \cite{OwenPRESENT}, from which this methodology is adapted, also illustrates eight leakage areas in their correlation power graphs. The difference is that all eight correlation troughs in that work have similar lengths, whereas the graphs in this work show different heights in the correlation branches. This discrepancy can be attributed to the effects of background noise interference. Therefore, frequency filtering was applied to the raw data to check for performance enhancement. Fortunately, improved results were obtained for the 45.08 MHz component when applied individually and together with other identified frequency elements. However, filtering at 56.33 MHz individually did not seem to enhance the results.

Being able to derive 80\% of the encryption key significantly reduces the time required for Brute Force Analysis (BFA). Once 8 bytes are retrieved, the remaining 2 bytes can be determined by applying a BFA as described in \cite{OwenPRESENT}. According to the SR results, it appears that both the number of ones in the key bytes and the placement of the bytes may impact the leakage accuracy.

Benchmarking was conducted to determine if any form of EM side-channel study has been able to retrieve the entire key of PRESENT. The only available study, a DEMA of PRESENT \cite{DEMA_PRESENT}, also managed to derive 8 bytes of the encryption key using over 12,000 waveforms, even with the use of side-channel evaluation hardware. Therefore, this work has captured the maximum possible number of key bits using the fewest EM waveforms with reference to the existing literature.

The idea of increasing the number of EM waveforms per attack for optimisation was inspired by other EM side-channel studies \cite{DEMA_PRESENT,cema_twine,cema_prince}, which utilised more than 10,000 waveforms touse the same methodology using different hardware such as Field Programmable Gate Arrays (FPGAs), other coefficient types, and frequency domains. Unlike using 256 or 2,048 waveforms, collecting over 10,000 waveforms can take several days to complete.

Unlike white-box attempts, it is challenging for an attacker to detect the intended operation, e.g., S-box, pLayer, etc., in black-box scenarios when the encryption processes through all 31 rounds. The difficulty of a successful attack increases if the manufacturer has implemented adequate countermeasures against potential EM leakages. Additionally, completely eliminating EM interferences is impossible. Therefore, it can be assumed that the full version of the PRESENT cipher in real-world applications would be sufficiently robust against EM side-channel leakages. It is expected to conduct the same methodology using different hardware such as Field Programmable Gate Arrays (FPGAs), other coefficient types and in the frequency domain in the future to evaluate the strength of PRESENT further.


\section{Conclusions}


Side-channel attack resilience is crucial in practical cryptanalysis to validate the expected security before actual deployment on devices. Therefore, continual attention must be given to side-channel analysis of cryptographic algorithms.  Only a few studies are available on EMA. This research examines the firmware robustness of PRESENT lightweight cipher against CEMA. The study began with a SEMA and a SEMFA, and then progressed to the CEMA. This attack model uses 256 waveforms per attack, which is the minimum number of EM traces for an EM attack compared to other EM attacks in the literature. The results indicate the possibility of deriving 8 bytes of the encryption key out of 10 bytes, substantially reducing the time required for BFAs.

The optimisation of the attack using 2048 waveforms further verifies the above performance by offering more stabilised results. At present, eight bytes of the PRESENT cipher are the maximum number of bytes derivable through physical attacks, according to both the literature and the results of this attack. To conclude, PRESENT cipher appears to satisfy the expected security when the full-round version is used, as opposed to reduced-round versions.

\bibliographystyle{IEEEtran}
\bibliography{bib}

\begin{thebibliography}{10}
\providecommand{\url}[1]{#1}
\csname url@samestyle\endcsname
\providecommand{\newblock}{\relax}
\providecommand{\bibinfo}[2]{#2}
\providecommand{\BIBentrySTDinterwordspacing}{\spaceskip=0pt\relax}
\providecommand{\BIBentryALTinterwordstretchfactor}{4}
\providecommand{\BIBentryALTinterwordspacing}{\spaceskip=\fontdimen2\font plus
\BIBentryALTinterwordstretchfactor\fontdimen3\font minus \fontdimen4\font\relax}
\providecommand{\BIBforeignlanguage}[2]{{%
\expandafter\ifx\csname l@#1\endcsname\relax
\typeout{** WARNING: IEEEtran.bst: No hyphenation pattern has been}%
\typeout{** loaded for the language `#1'. Using the pattern for}%
\typeout{** the default language instead.}%
\else
\language=\csname l@#1\endcsname
\fi
#2}}
\providecommand{\BIBdecl}{\relax}
\BIBdecl

\bibitem{PRESENT}
A.~Bogdanov, L.~R. Knudsen, G.~Leander, C.~Paar, A.~Poschmann, M.~J.~B. Robshaw, Y.~Seurin, and C.~Vikkelsoe, ``{PRESENT: An Ultra-Lightweight Block Cipher},'' in \emph{Cryptographic Hardware and Embedded Systems (CHES)}, P.~Paillier and I.~Verbauwhede, Eds.\hskip 1em plus 0.5em minus 0.4em\relax Springer, Berlin, Heidelberg, 2007, pp. 450--466, doi: \url{10.1007/978-3-540-74735-2\_31}.

\bibitem{ISOIEC29167}
\BIBentryALTinterwordspacing
{ISO/IEC ICS 35.030 IT Security}, ``{Information Security — Lightweight Cryptography},'' \emph{ISO/IEC JTC 1/SC 27 Information security, Cybersecurity and Privacy Protection}, Nov. 2019, (Date accessed: Jan. 21, 2020). [Online]. Available: \url{https://www.iso.org/standard/78477.html}
\BIBentrySTDinterwordspacing

\bibitem{NISTIR8114}
K.~A. McKay, L.~Bassham, M.~S. Turan, and N.~Mouha, ``{Report on Lightweight Cryptography},'' \emph{National Institute of Standards and Technology}, Mar. 2017, doi: \url{10.6028/NIST.IR.8114}.

\bibitem{nilu22}
N.~A. Gunathilake, A.~Al-Dubai, W.~J. Buchanan, and O.~Lo, ``{Electromagnetic Side-Channel Attack Resilience against PRESENT Lightweight Block Cipher},'' in \emph{6th International Conference on Cryptography, Security and Privacy (CSP)}, 2022, pp. 51--55, doi: \url{10.1109/CSP55486.2022.00018}.

\bibitem{DEMA_PRESENT}
Y.~Nozaki, T.~Iwase, Y.~Ikezaki, and M.~Yoshikawa, ``{Differential Electromagnetic Analysis for PRESENT and its Evaluation with Several Selection Functions},'' \emph{Journal of International Council on Electrical Engineering}, vol.~7, no.~1, pp. 137--141, Jun. 2017.

\bibitem{cema_twine}
M.~Yoshikawa, Y.~Nozaki, and K.~Asahi, ``{Electromagnetic Analysis Attack for a Lightweight Block Cipher TWINE},'' in \emph{IEEE/ACES International Conference on Wireless Information Technology and Systems (ICWITS) and Applied Computational Electromagnetics (ACES)}, 2016, pp. 1--2, doi: \url{10.1109/ROPACES.2016.7465354}.

\bibitem{EM_freq2}
Q.~Le, L.~Miralles, A.~Sayakkara, N.-A. Le-Khac, and M.~Scanlon, ``{Identifying Internet of Things Software Activities using Deep Learning-Based Electromagnetic Side-Channel Analysis},'' \emph{Forensic Science International: Digital Investigation}, vol.~39, Oct. 2021, doi: \url{10.1016/j.fsidi.2021.301308}.

\bibitem{SCA_freq_IoT}
A.~Sayakkara, N.-A. Le-Khac, and M.~Scanlon, ``{Leveraging Electromagnetic Side-Channel Analysis for the Investigation of IoT Devices},'' \emph{Digital Investigation}, vol.~29, pp. S94--S103, Jul. 2019, doi: \url{10.1016/j.diin.2019.04.012}.

\bibitem{F_SCA_AES}
C.~C. Tiu, ``{A New Frequency-Based Side-Channel Attack for Embedded Systems},'' Electrical and Computer Engineering, University of Waterloo, Waterloo, Ontario, Canada, Tech. Rep., 2005, mSc thesis.

\bibitem{SEMA}
\BIBentryALTinterwordspacing
A.~Lakshminarasimhan, \emph{{Electromagnetic Side-Channel Analysis for Hardware and Software Watermarking}}, MSc dissertation, Department of Electrical and Computer Engineering, University of Massachusetts Amherst, Sep. 2011, (Date accessed: May 28, 2019). [Online]. Available: \url{https://scholarworks.umass.edu/theses/693}
\BIBentrySTDinterwordspacing

\bibitem{simple_tem_EMA}
Z.~Martinasek, V.~Zeman, and K.~Trasy, ``{Simple Electromagnetic Analysis in Cryptography},'' \emph{International Journal of Advances in Telecommunications, Electrotechnics, Signals and Systems}, vol.~1, Jul. 2012, doi: \url{10.11601/ijates.v1i1.6}.

\bibitem{TemplateSCA}
S.~Chari, J.~R. Rao, and P.~Rohatgi, ``{Template Attacks},'' in \emph{Cryptographic Hardware and Embedded Systems (CHES)}, B.~S. Kaliski, K.~Ko\c{c}, and C.~Paar, Eds., vol. 2523.\hskip 1em plus 0.5em minus 0.4em\relax Springer, Berlin, Heidelberg, Feb. 2002, doi: \url{10.1007/3-540-36400-5\_3}.

\bibitem{Sayakkara_2019}
A.~Sayakkara, N.-A. Le-Khac, and M.~Scanlon, ``{A Survey of Electromagnetic Side-channel Attacks and Discussion on their Case-Progressing Potential for Digital Forensics},'' \emph{Digital Investigation}, vol.~29, p. 43–54, Jun. 2019, doi: \url{10.1016/j.diin.2019.03.002}.

\bibitem{localised_gini}
J.~Heyszl, S.~Mangard, B.~Heinz, F.~Stumpf, and G.~Sigl, ``{Localised Electromagnetic Analysis of Cryptographic Implementations},'' vol. 7178, Jan. 2012, pp. 231--244, doi: \url{10.1007/978-3-642-27954-6\_15}.

\bibitem{robyns}
\BIBentryALTinterwordspacing
P.~Robyns. (2019, Feb.) {Performing Low-cost Electromagnetic Side-Channel Attacks using RTL-SDR and Neural Networks}. {FOSDEM}. Brussels, Belgium. (Date accessed: Apr. 10, 2020). [Online]. Available: \url{https://youtu.be/cs08QSIbp-A}
\BIBentrySTDinterwordspacing

\bibitem{OwenPRESENT}
O.~Lo, W.~J. Buchanan, and D.~Carson, ``{Correlation Power Analysis on the PRESENT Block Cipher on an Embedded Device},'' in \emph{13th International Conference on Availability, Reliability and Security}.\hskip 1em plus 0.5em minus 0.4em\relax Hamburg, Germany: Association for Computing Machinery, 2018, doi: \url{10.1145/3230833.3232801}.

\bibitem{cema_prince}
M.~Yoshikawa and Y.~Nozaki, ``{Electromagnetic Analysis Attack for a Lightweight Cipher PRINCE},'' in \emph{IEEE International Conference on Cybercrime and Computer Forensic (ICCCF)}, Vancouver, Canada, 2016, pp. 1--6, doi: \url{10.1109/ICCCF.2016.7740423}.

\end{thebibliography}

\end{document}